\documentclass[twocolumn,superscriptaddress,showpacs,preprintnumbers,amsmath,amssymb]{revtex4}
\usepackage{amsmath}
\usepackage{amssymb}
\usepackage{dcolumn}
\usepackage{graphicx}
\usepackage{bm}
\usepackage{epstopdf}
\usepackage{threeparttable}
\usepackage{float}
\usepackage[export]{adjustbox}

\begin{document}
\title{On the origin of multi-component bulk metallic glasses: Atomic size mismatches and de-mixing}
\author{Kai Zhang} 
\affiliation{Department of Mechanical Engineering and Materials Science, Yale University, New Haven, Connecticut, 06520, USA}
\affiliation{Center for Research on Interface Structures and Phenomena, Yale University, New Haven, Connecticut, 06520, USA}

\author{Bradley Dice} 
\affiliation{William Jewell College, Liberty,  Missouri, 64068, USA}
\affiliation{Center for Research on Interface Structures and Phenomena, Yale University, New Haven, Connecticut, 06520, USA}

\author{Yanhui Liu} 
\affiliation{Department of Mechanical Engineering and Materials Science, Yale University, New Haven, Connecticut, 06520, USA}
\affiliation{Center for Research on Interface Structures and Phenomena, Yale University, New Haven, Connecticut, 06520, USA}
\author{Jan Schroers}
\affiliation{Department of Mechanical Engineering and Materials Science, Yale University, New Haven, Connecticut, 06520, USA}
\affiliation{Center for Research on Interface Structures and Phenomena, Yale University, New Haven, Connecticut, 06520, USA}
\author{Mark D. Shattuck}
\affiliation{Department of Physics and Benjamin Levich Institute, The City College of the City University of New York, New York, New York, 10031, USA}
\affiliation{Department of Mechanical Engineering and Materials Science, Yale University, New Haven, Connecticut, 06520, USA}
\author{Corey S. O'Hern}
\affiliation{Department of Mechanical Engineering and Materials Science, Yale University, New Haven, Connecticut, 06520, USA}
\affiliation{Center for Research on Interface Structures and Phenomena, Yale University, New Haven, Connecticut, 06520, USA}
\affiliation{Department of Physics, Yale University, New Haven, Connecticut, 06520, USA}
\affiliation{Department of Applied Physics, Yale University, New Haven, Connecticut, 06520, USA}

\date{\today}

\begin{abstract}
The likelihood that an undercooled liquid vitrifies or crystallizes
depends on the cooling rate $\mathcal R$. The critical cooling rate $\mathcal{R}_c$,
below which the liquid crystallizes upon cooling, characterizes the
glass-forming ability (GFA) of the system. While pure metals are
typically poor glass formers with $\mathcal {R}_c>10^{12}~{\rm K/s}$, specific
multi-component alloys can form bulk metallic glasses (BMGs) even at
cooling rates below $\mathcal {R}\sim 1~{\rm K/s}$. Conventional wisdom asserts
that metal alloys with three or more components are better glass
formers (with smaller ${\cal R}_c$) than binary alloys. However, there is
currently no theoretical framework that provides quantitative
predictions for $\mathcal{R}_c$ for multi-component alloys.  In this manuscript,
we perform simulations of ternary hard-sphere systems, which have been
shown to be accurate models for the glass-forming ability of BMGs, to
understand the roles of geometric frustration and demixing in
determining $\mathcal {R}_c$. Specifically, we compress ternary hard sphere
mixtures into jammed packings and measure the critical compression rate,
below which the system crystallizes, as a function of the diameter
ratios $\sigma_B/\sigma_A$ and $\sigma_C/\sigma_A$ and number
fractions $x_A$, $x_B$, and $x_C$.  We find two distinct regimes for the GFA
in parameter space for ternary hard spheres. When the diameter ratios
are close to $1$, such that the largest ($A$) and smallest
($C$) species are well-mixed, the GFA of ternary systems is no
better than that of the optimal binary glass former. However, when
$\sigma_C/\sigma_A \lesssim 0.8$ is below the demixing threshold for
binary systems, adding a third component $B$ with $\sigma_C < \sigma_B
< \sigma_A$ increases the GFA of the system by preventing demixing of
$A$ and $C$. Analysis of the available data from experimental studies
indicates that most ternary BMGs are below the binary demixing
threshold with $\sigma_C/\sigma_A < 0.8$.

\end{abstract}

 \maketitle

\section{Introduction} 
\label{intro}

When atomic and molecular liquids are cooled sufficiently rapidly
({\it i.e.} above the critical cooling rate $\mathcal{R}_c$), they
bypass crystallization and become trapped in disordered glassy
configurations~\cite{debenedetti:2001}. Avoiding crystallization in
pure metals is very challenging and has only been achieved in
experiments recently~\cite{zhong:2014}. On the other hand,
multi-component liquid alloys can form bulk metallic glasses (BMGs)
that possess centimeter or greater casting thicknesses and critical
cooling rates $\mathcal{R}_c < 1~{\rm
  K/s}$~\cite{chen,greer:1995,inoue:2000}.  BMGs have shown great
promise as structural materials because they are amorphous with few
defects and possess higher processability than crystalline
metals~\cite{greer:2007,schroers}.

The conventional wisdom in the BMG research community is that BMGs
should contain three or more atomic species~\cite{greer:2009} with
atomic size differences above 12\% ({\it i.e.} the ratio of the
diameters of the smallest to the largest species should be $\lesssim
0.89$)~\cite{inoue}.  Intuitively, more atomic components with
different sizes introduces geometric frustration or ``confusion,''
which delays crystallization~\cite{greer:1993,greer:1995,zhang:1991}.
Also, it has been suggested that a mixture of multiple atomic species
leads to dense packing in the liquid state and thus enhanced stability
of the glass~\cite{zhang:1991}. The minimum critical cooling rate
observed for binary BMGs is $\mathcal{R}_c \sim 10^2~{\rm K/s}$, while
it decreases to $10^{-1}$ and $10^{-2}~{\rm K/s}$ for ternary and
quaternary systems, respectively.  Alloys with similarly sized atomic
constituents can only be cast into glassy thin films. (See
Table~\ref{table:gfa} in Appendix~\ref{app1}.) Even with these
empirical rules, there is an enormous parameter space of potential
BMGs and we lack a complete theoretical framework that would enable the
prediction of $\mathcal{R}_c$ for each alloy in the design space.

A number of recent studies of hard-sphere mixtures have shown that
dense atomic packing plays an important role in determining the
glass-forming ability (GFA) of metallic
alloys~\cite{egami:1984,miracle:2003,miracle:2004,jalali:2004,sheng:2006,miracle:2013}. In
particular, binary metal-metal ({\it i.e.} transition metal-transition
metal) BMGs, such as Cu-Zr, Cu-Hf, and Ca-Al, possess atomic size ratios
$\alpha=\sigma_B/\sigma_A$ and compositions $x_B$ that occur in the
region of parameter space with the smallest $\mathcal{R}_c$ for binary hard
spheres~\cite{zhang:2014}. 

\begin{figure}
\includegraphics[width=3.5in]{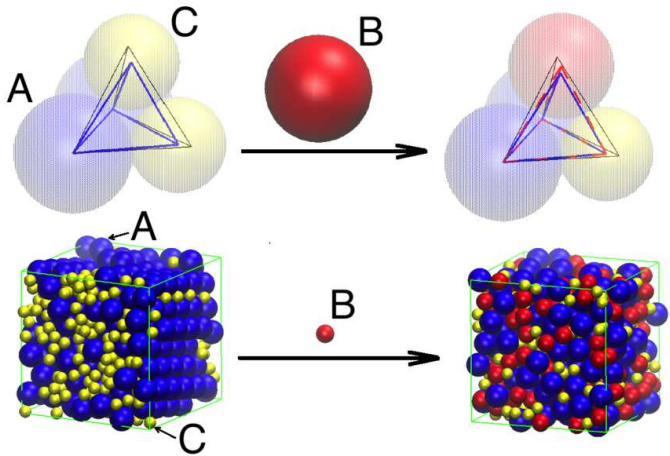}
\caption{(Color online) (top) Comparison of the volumes of tetrahedral
cells formed by the centers of face-centered cubic packed atoms with
(left) two versus (right) three different sizes ({\it i.e.} either 
$\sigma_C/\sigma_A=0.8$ with $x_B = 0$ or 
$\sigma_B/\sigma_A=0.9$, $\sigma_C/\sigma_A=0.8$, and $x_B>0$). 
The tetrahedral volume for four
same-sized atoms is $0.118~\sigma_A^3$ (thin black line), while it
is $0.084~\sigma_A^3$ for two large and two small atoms (thick blue
solid line) and $0.092~\sigma_A^3$ for one small, one
intermediate, and two large sized atoms (red dashed line).  Thus, 
the distortion of
the tetrahedral cell is smaller for the ternary system. (bottom) When
atoms of an intermediate size ($\sigma_B/\sigma_A = 0.75$) are added to a
binary system (with diameter ratio $\sigma_C/\sigma_A = 0.5$), it becomes more uniformly mixed and less ordered, {\it i.e.} with global bond 
orientational order parameter
$Q_6=0.02$ (right) compared to $0.15$ (left).}
\label{fig:twocases}
\end{figure}

Can hard sphere models accurately capture the dependence of the GFA on
the atomic size ratios and compositions for {\it multi-component}
alloys? In this manuscript, we study the glass-forming ability of
ternary hard-sphere mixtures. We find two key results: (1) When the
sizes of the three components are comparable, ternary systems behave
similar to binary systems, and the GFA cannot be larger than that of a
binary system consisting of the largest and smallest components. In
this case, the packing efficiency of the ternary system is close to
that of the binary systems. (See Fig.~\ref{fig:twocases} (top).) (2)
When the diameter ratio of the smallest to the largest component is
beyond the demixing limit ($\alpha\lesssim 0.8$~\cite{zhang:2014}), adding a
third component with an intermediate size can increase the GFA by
preventing demixing. In this scenario, the packing fraction of the
ternary system is significantly higher than the demixed binary
system. (See Fig.~\ref{fig:twocases} (bottom).)  This demixing mechanism has
also been found in studies of segregation of granular media and other
particulate solids~\cite{olsen:1964}.

\section{Methods} 
\label{methods}

We performed event-driven molecular dynamics simulations of $N=500$
ternary hard spheres with diameters $\sigma_A \ge \sigma_B \ge
\sigma_C$, number fractions $x_A = N_A/N$, $x_B = N_B/N$, and $x_C =
N_C/N$, and the same mass $m$.  We compressed systems initially
prepared in liquid states at packing fraction $\phi = 0.25$ so that
they exponentially approach static jammed packings at $\phi=\phi_J$ as
a function of time.  We terminate each compression run when
$(\phi_J-\phi)/\phi_J = 10^{-3}$. We vary the compression rate $R$
over $5$ orders of magnitude~\cite{zhang:2014}. Note that $R$ is given
in units of $\sqrt{k_B T/m\sigma_A^2}$ and $R=1$ corresponds to a
cooling rate $\mathcal{R}\approx 10^{12}~{\rm K/s}$ for metal
alloys~\cite{truskett:2000}.  The crystal structures that compete with
glass formation possess face-centered cubic (FCC)-like order, and thus
we characterize the positional order of the packings using the global
bond orientational order parameter $Q_6$~\cite{steinhardt:1983}
averaged over $96$ independent compression runs. The critical
compression rate $R_c$ is determined by the intersection of the mean
and median $Q_6$ as a function of $R$. (See Appendix~\ref{app2}.) To
explore the glass-forming ability diagram for ternary systems, we
studied more than $20$ compositions and $10$ pairs of atomic size
ratios $\sigma_B/\sigma_A$ and $\sigma_C/\sigma_A$. Additional details
of the simulation methods can be found in Ref.~\cite{zhang:2014}.

\section{Results}
\label{results}
 
\begin{figure}
\includegraphics[width=3.5in]{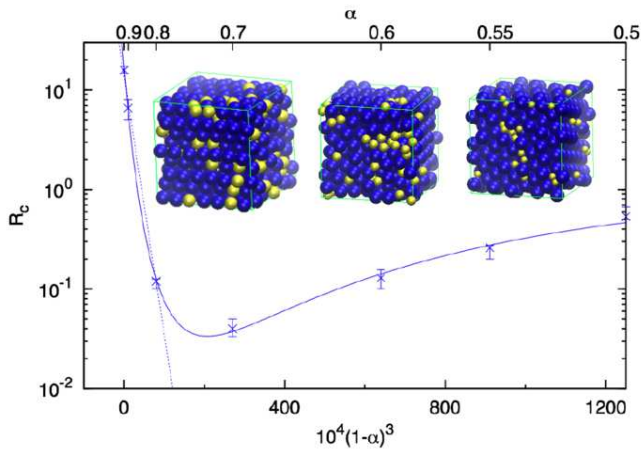}
\caption{(Color online) In binary hard-sphere mixtures with diameter
ratio $\alpha > \alpha_c(x_B)$ and fixed number fraction $x_B$ of small particles, 
$\log_{10} R_c$ drops linearly with
$(1-\alpha)^3$ (dotted line).  
For $\alpha < \alpha_c$, the large and small particles demix and
$R_c$ begins to increase with decreasing $\alpha$.  The composition-dependent threshold is  $\alpha_c \approx 0.8$ for $x_B = 0.2$. The inset shows
snapshots of configurations at $R\sim R_c$ for $x_B = 0.2$ and
$\alpha=0.9$, $0.7$, and $0.5$ from left to right, which illustrates
increasing demixing as $\alpha$ decreases. }
\label{fig:Rc_alpha}
\end{figure}

In our previous studies of binary hard-sphere mixtures with diameter
ratio $\alpha=\sigma_B/\sigma_A$ and small particle number fraction
$x_B$, we found that the critical compression rate $R_c$ decreases
exponentially, $R_c \sim \exp[C(x_B)(1-\alpha)^3]$, where $C$ is a
composition-dependent constant, for $\alpha \gtrsim 0.8$ above the
demixing limit.  (See Fig.~\ref{fig:Rc_alpha}.) In contrast, for
$\alpha \lesssim 0.8$ the large and small particles in binary systems
can demix, which then induces crystallization. Thus, the glass-forming ability
for binary hard-sphere systems first increases with decreasing
$\alpha$, but then begins to decrease for $\alpha \lesssim
0.8$~\cite{zhang:2014}.

\begin{figure}
\includegraphics[width=3.5in]{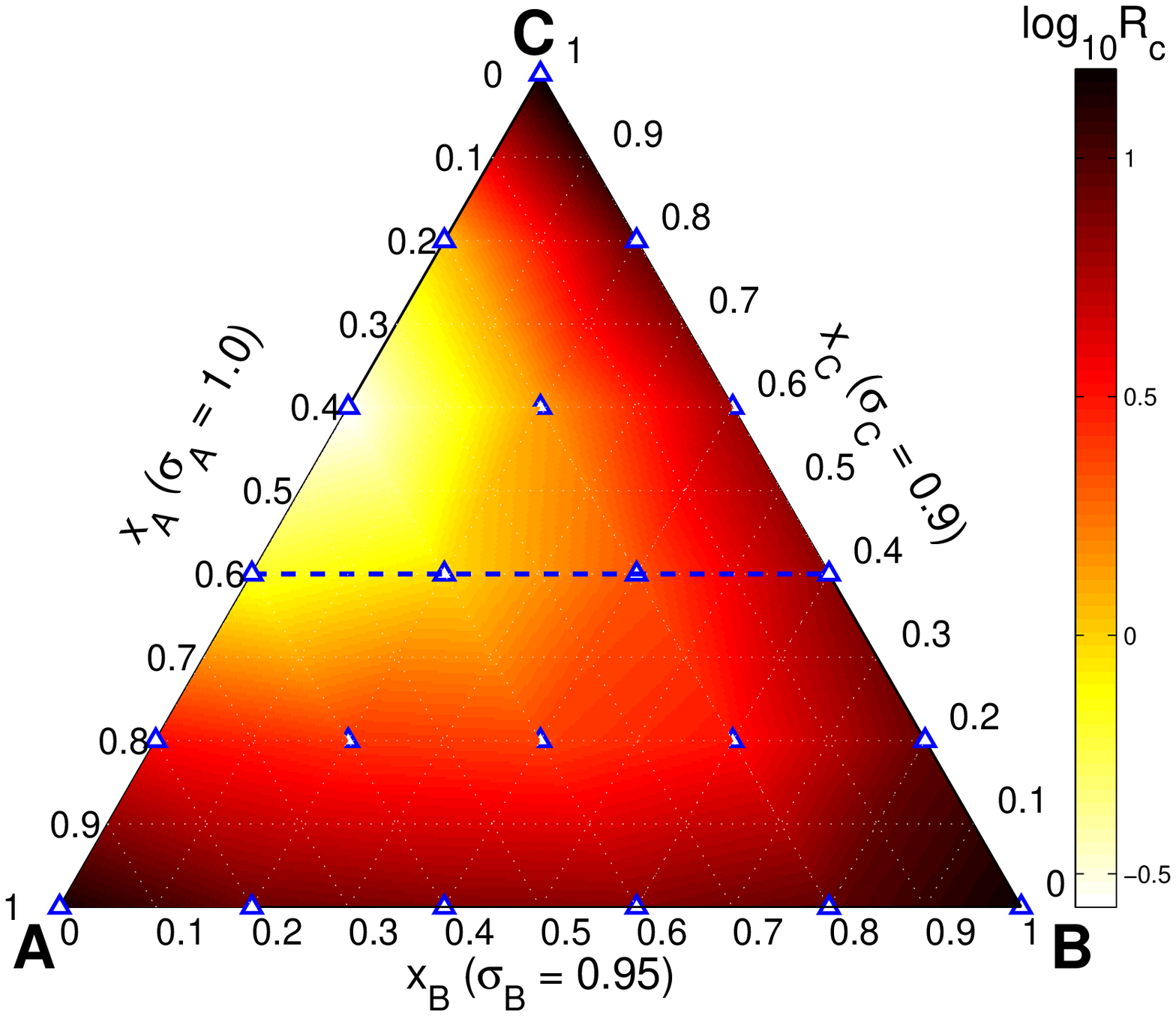}
\includegraphics[width=3.5in]{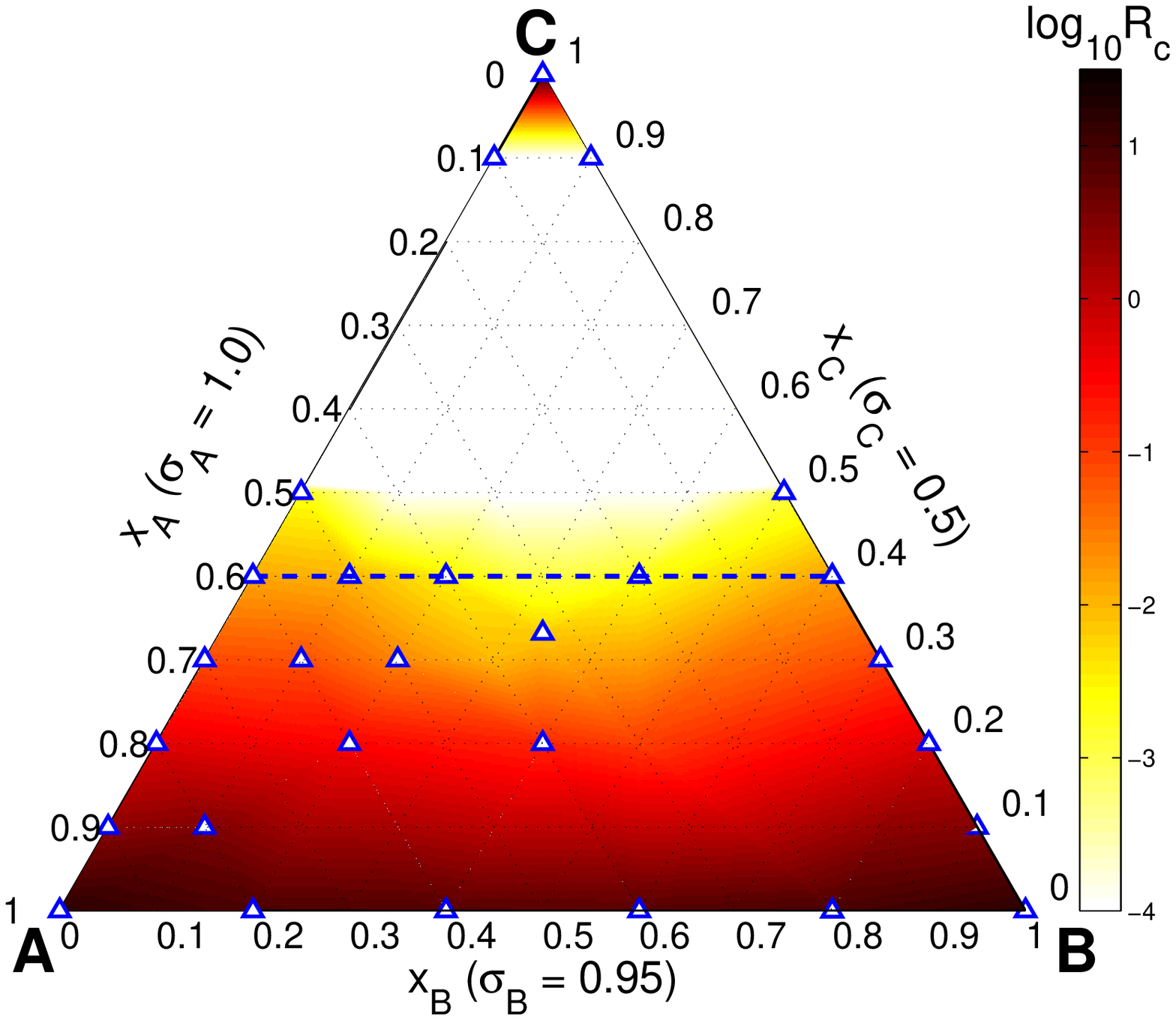}
\includegraphics[width=3in]{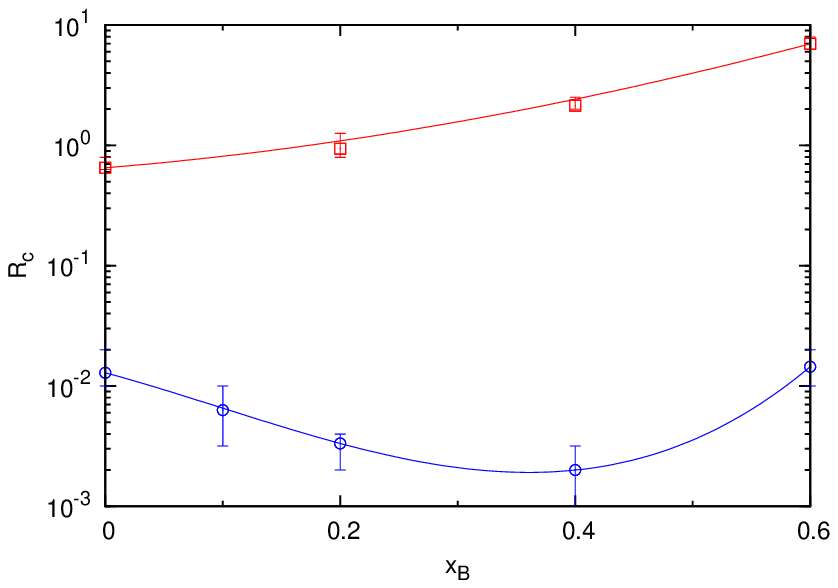}
\caption{(Color online) (top) Critical compression rate $R_c$ as a
function of the compositions $x_A$, $x_B$, and $x_C$ in ternary
hard-sphere systems with diameter ratios $\sigma_B/\sigma_A=0.95$
and $\sigma_C/\sigma_A=0.9$. The minimum $R_c$ occurs on the 
edge $\overline{AC}$, {\it i.e.} binary systems with $x_B=0$. Moving 
perpendicular to $\overline{AC}$ on the diagram causes increases in $R_c$.
(middle) Same as the top plot, but with diameter ratios 
$\sigma_B/\sigma_A=0.95$ and $\sigma_C/\sigma_A=0.5$. The minimum 
$R_c$ no longer occurs for binary systems on the edge 
$\overline{AC}$.  Note that the contour plots of $\log_{10} R_c$ are 
interpolated from $\sim 20$ simulation runs (triangles) and given on 
a color scale that decreases from dark to light. (bottom) $R_c$ as a function of $x_B$ at fixed $x_C=0.4$ (dashed line in top and middle panels) for the diameter ratios studied in the top (squares) and middle (circles) panels.}
\label{fig:Rc_x}
\end{figure}

We first focus on ternary hard-sphere systems with weak size
disparities. In Fig.~\ref{fig:Rc_x} (top), we plot the critical
compression rate $R_c$ as a function of the number fractions of the
three components $x_A$, $x_B$ and $x_C$ at fixed diameter ratios
$\sigma_B/\sigma_A=0.95$ and $\sigma_C/\sigma_A=0.9$.  We find that
the best glass-forming regime ({\it i.e.} the region with the smallest
$R_c$) occurs on the binary composition line $\overline{AC}$ near
$x_C=1-x_A \approx 0.6$ and $x_B=0$. Adding the third component $B$
with an intermediate size $\sigma_C<\sigma_B<\sigma_A$ causes a
decrease in the glass-forming ability (or increase in $R_c$). Since
the competing crystal for bidisperse hard-sphere systems in this
diameter ratio regime is a deformed FCC crystal, adding a third
component with an intermediate size to a binary system reduces the
lattice distortion (Fig.~\ref{fig:twocases} (top)) {\it and} the
glass-forming ability.  These results are consistent with
experimental observations for bulk metallic glasses, which are
summarized in Table~\ref{table:gfa} in Appendix~\ref{app1}.  There are
no observed ternary bulk metallic glasses with weak size
polydispersity, {\it i.e.} the diameter ratio of the smallest to the
largest component satisfies $\alpha \gtrsim 0.8$.
    
We now consider ternary hard-sphere systems with a diameter
ratio disparity that is beyond the demixing threshold, {\it i.e.}
$\sigma_C/\sigma_A<0.8$. In Fig.~\ref{fig:Rc_x} (middle), we plot the
critical compression rate $R_c$ as a function of the compositions
$x_A$, $x_B$, and $x_C$ for ternary systems with diameter ratios
$\sigma_B/\sigma_A=0.95$ and $\sigma_C/\sigma_A=0.5$. For this system
the smallest value of $R_c$ does not occur for $x_B = 0$.  In contrast, 
we show that $R_c$ decreases with increasing $x_B$.  (See Fig.~\ref{fig:Rc_x}
(bottom).) 

\begin{figure}
\includegraphics[width=3.5in]{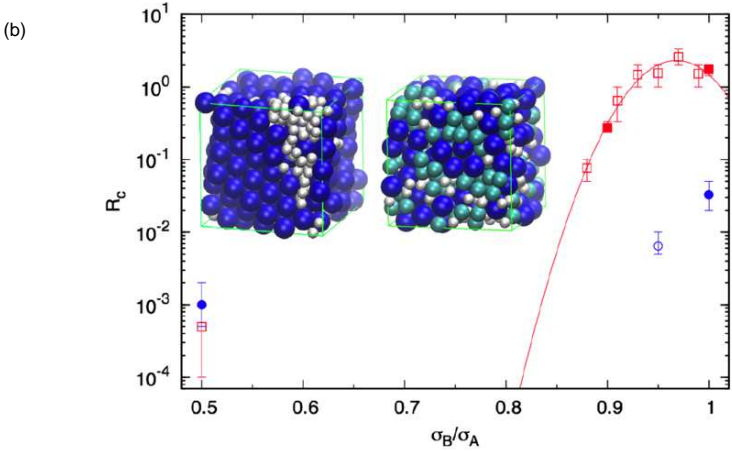}
\caption{(Color online) The critical compression rate $R_c$ as a 
function of the diameter ratio $\sigma_B/\sigma_A$ at fixed composition 
$x_A=x_B=x_C=1/3$ and $\sigma_C/\sigma_A=0.9$ 
(squares) or $\sigma_C/\sigma_A=0.5$ (circles). The solid lines 
are polynomial fits to the data to show qualitative trends. Ternary systems 
reduce to binary systems
when $\sigma_B=\sigma_A$ or $\sigma_C$ (solid symbols). In the inset, 
we show configurations obtained at a slow compression rate $R =10^{-3}$ for
$\sigma_B/\sigma_A=0.5$ (left) and $0.75$ (right). Large, intermediate, 
and small particles are shaded from dark to light. }
\label{fig:Rc_sigma}
\end{figure}

We can also measure the glass-forming ability at fixed composition and
vary the diameters of one of the particles.  In
Fig.~\ref{fig:Rc_sigma}, we fix the compositions $x_A=x_B=x_C=1/3$ and
diameters $\sigma_A$ and $\sigma_C$ of two components and measure
$R_c$ as a function of the diameter ratio $\sigma_B/\sigma_A$. Note
that, when $\sigma_B=\sigma_A$ ($\sigma_C$), the ternary systems
reduce to binary systems with $x_C=1/3$ ($2/3$). In experimentally
observed ternary BMGs, when the diameters of two of the three
components are similar, for instance CuNi and AlTi, the ternary
glass-forming ability diagram is symmetric and equivalent to that of
the corresponding binary system~\cite{takeuchi:2001}.  We first focus
on ternary systems with $\sigma_C/\sigma_A = 0.9$, which does not lead
to demixing. When $\sigma_C/\sigma_A<\sigma_B/\sigma_A<1$, $R_c$ has a
maximum at $\sigma_B/\sigma_A < 1$ and the ternary systems are worse
glass formers than binary systems with $\sigma_B=\sigma_C$. These
ternary systems show enhanced glass-forming ability above that for
binary systems only when $\sigma_B/\sigma_A \lesssim 0.9$. (See
Fig.~\ref{fig:Rc_sigma}.)

In Fig.~\ref{fig:Rc_sigma}, we also consider fixed diameter ratio
$\sigma_C/\sigma_A=0.5$ for which the two components tend to demix. In
this case, $R_c$ does not possess a maximum at $\sigma_B/\sigma_A <
1$, and thus these ternary systems can possess enhanced glass-forming
ability compared to corresponding binary systems. The introduction of
the third component with an intermediate size $\sigma_B$ prevents
demixing. As shown in the insets of Fig.~\ref{fig:Rc_sigma}, binary
systems with $\sigma_B/\sigma_A=0.5$ demix and crystallize (left),
while ternary systems with $\sigma_B/\sigma_A=0.75$ remain well-mixed and
amorphous (right). Although the large particles $A$ exclude the small ones
$C$, $A$ particles mix with $B$ particles and $B$ particles mix with
$C$ particles, which leads to effective mixing of $A$ and $C$ particles.

Because packing efficiency and vibrational entropy determine the
stability of crystals in hard-sphere
systems~\cite{turnbull:1961,parisi:2010}, one can correlate the
packing fraction at jamming $\phi_J$ with the critical compression
rate $R_c$ as demonstrated in binary systems~\cite{zhang:2014}. We
study three relevant packing fractions: $\phi_J^{\rm RCP}$ obtained
in the limit $R \rightarrow
\infty$~\cite{torquato:2000,torquato:2010}, $\phi_J^a$ for amorphous
packings obtained at $R \sim R_c$, and $\phi_J^x$ for partially crystalline
packings obtained at $R \sim R_c$. We find that the packing fraction
of single-crystal FCC packings is not strongly correlated with $R_c$. 
Instead, partially crystalline systems compete with
glass formation~\cite{zhang:2014}.  

\begin{figure}
\includegraphics[width=3.5in]{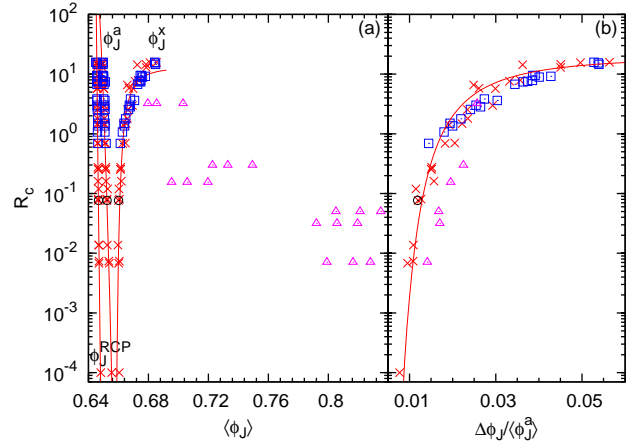}
\caption{(Color online) (a) For each binary and ternary system, we plot
the corresponding critical compression rate $R_c$ and three definitions 
of the packing
fraction at jamming: $\phi_J^{\rm RCP}$ obtained in the $R \rightarrow
\infty$ limit, $\phi_J^a$ for amorphous packings obtained at $R\sim R_c$,
and $\phi_J^x$ for partially crystalline packings obtained at $R\sim R_c$ with
$\phi_J^{\rm RCP} <\phi_J^a<\phi_J^x$. 
The solid lines give polynomial fits to the data for the packing 
fraction at jamming for $\alpha \ge 0.8$.
(b) $R_c$ 
plotted versus the normalized difference $\Delta \phi_J/\langle 
\phi_J^a \rangle$, where $\Delta \phi_J = \phi_J^x -\phi_J^a$. The master curve (solid 
line) obeys $\log_{10} R_c \sim (\Delta \phi_J/\langle \phi_J^a\rangle)^{-2}$~\cite{zhang:2014}. In (a) and 
(b), we considered binary systems with diameter ratios $\alpha \ge 0.8$ 
(crosses) and ternary systems with diameter ratio pairs $\sigma_B/\sigma_A = 0.95$ 
and $\sigma_C/\sigma_A = 0.9$, $\sigma_B/\sigma_A =0.9$ and $\sigma_C/\sigma_A = 0.88$, and $\sigma_B/\sigma_A = 0.95$ and $\sigma_C/\sigma_A = 0.5$ 
(squares, circles, and triangles, respectively). In the absence of demixing, 
$R_c$ versus the jammed packing fraction for ternary systems is 
quantitatively similar to that for binary 
systems. However, ternary systems that demix deviate
from the master curve.}
\label{fig:Rc_phi}
\end{figure}

As shown in Fig.~\ref{fig:Rc_phi} for ternary systems with diameter
ratio pairs $\sigma_B/\sigma_A=0.95$ and $\sigma_C/\sigma_A=0.9$ and
$\sigma_B/\sigma_A = 0.9$ and $\sigma_C/\sigma_A=0.88$ that do not
demix, the relations between $R_c$ and packing fraction $\phi_J$ at
jamming follow the trends for binary systems, {\it i.e.} as $\phi_J^a$
and $\phi_J^x$ approach each other, $R_c \rightarrow 0$.  In addition,
the packing fraction in these ternary systems is not larger than in
binary systems, contrary to the intuition that ternary systems are
always denser than binary systems and thus possess higher
glass-forming abilities~\cite{zhang:1991}. These results show that
ternary systems with weak diameter ratio disparities can be described
effectively as binary systems since the additional intermediate-sized
particles only decreases the particle size gradient in the original
binary system without changing the mechanism that drives
crystallization~\cite{li:2008,zhang:2014}.  The deviations from the
master curve (solid lines) in Fig.~\ref{fig:Rc_phi}
indicate demixing in the ternary system with diameter ratio pairs
$\sigma_B/\sigma_A = 0.95$ and $\sigma_C/\sigma_A = 0.5$, and in this
case higher packing fractions are obtained than for binary systems.

\begin{figure}
\includegraphics[width=4.8in,center]{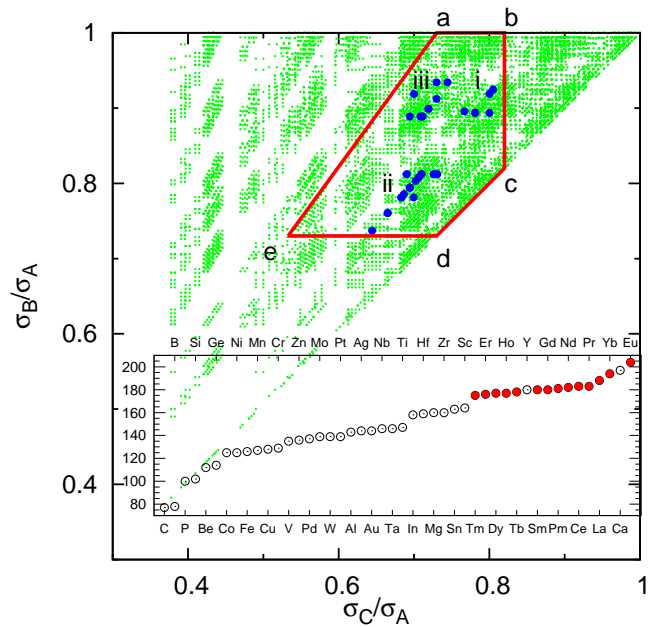}
\caption{(Color online) Scatter plot of the diameter ratios
$\sigma_B/\sigma_A$ versus $\sigma_C/\sigma_A$ with 
$\sigma_C\le\sigma_B\le\sigma_A$ for all $15180$ combinations of three elements chosen from
$46$ possible BMG-forming elements (dots). $33$ of the combinations (filled 
circles) have
been shown experimentally to form BMGs, and occur in roughly three 
main clusters: i) (Zr,Hf,Sn,Mg)-(Al,Ti,Nb)-(Cu,Ni,Co),
ii) (Y,Ln)-Al-(Cu,Ni,Co), and iii) (Au,Pd,Pt)-(Cu,Ni)-(Si,P).  The predicted
BMG-forming alloys are located within the polygon bounded by the solid 
lines.  The inset gives the atomic radii in ${\rm pm}$ of the $46$ 
potential BMG-forming elements ordered from smallest to 
largest~\cite{egami:1984,inoue:2000,miracle:2003}. The symbols of the elements
alternate from top to bottom on the horizontal axis and the lanthanide 
elements are shown as filled circles.}
\label{fig:experiment}
\end{figure}

In Fig.~\ref{fig:experiment}, we illustrate the diameter ratio
variation for $15180$ combinations of three elements from a set of
$46$ potential BMG-forming elements. The atomic sizes of the
elements are given by their metallic
radii~\cite{egami:1984,miracle:2003}, which are shown in the inset of
Fig.~\ref{fig:experiment}. We define the component types so that they
satisfy $\sigma_C\le\sigma_B\le\sigma_A$, and thus the data occurs in
left corner of the plot.

In previous studies, we predicted that the optimal binary hard-sphere
glass formers occur in the diameter ratio range $0.73 < \alpha <
0.82$, where $\alpha$ is sufficiently small to prevent ordering, but
not too small to cause demixing. We can estimate the optimal
glass-forming regime in the $\sigma_B/\sigma_A$ and
$\sigma_C/\sigma_A$ plane for ternary systems using the following
arguments.  First, two of the boundaries ($\overline{ab}$ and
$\overline{cd}$) can be obtained directly from the results for binary
systems.  Because adding a third intermediate-sized component can
prevent demixing of the original two components, we expect that the
lower bound for the diameter ratio in ternary BMG-forming systems to
be much smaller than $0.73$.  We propose that $0.73^2$ is the lower
bound for the diameter ratio for ternary systems. In this case
$\sigma_C/\sigma_A = 0.73^2$, $\sigma_B/\sigma_A = 0.73$, and
$\sigma_C/\sigma_B=0.73$ (point $e$ in Fig.~\ref{fig:experiment}), and
thus all binary combinations are above the lower bound of the good GFA regime.
 
We predict that good BMG-forming alloys will occur within the polygon
defined by lines connecting the points (a)-(e) in the
$\sigma_B/\sigma_A$ and $\sigma_C/\sigma_A$ parameter space.  $33$
ternary alloy systems have been observed experimentally in amorphous
states (filled circles)~\cite{inoue:2000,miracle:2003,long:2009}, all
of which fall in the good glass-forming regime predicted by
hard-sphere systems. For the experimentally observed ternary BMGs, the
diameter ratio for the smallest versus the largest particle
$\sigma_C/\sigma_A$ satisfies $\alpha<0.8$, which is below the
demixing limit for binary systems. Therefore, the experimentally
observed ternary BMGs have better GFA than the best binary
glass-forming alloys.  In addition, the experimentally observed BMGs
tend to be positioned away from the boundaries $\overline{ab}$ and
$\overline{cd}$. As ternary systems approach $\overline{ab}$
($\overline{cd}$), $\sigma_B/\sigma_A \rightarrow 1$
($\sigma_C/\sigma_A \rightarrow 1$), which causes them to behave
as binary systems and reduces their glass-forming ability. The 
experimentally observed ternary BMGs cluster roughly into three groups:
(i) (Zr,Hf,Sn,Mg)-(Al,Ti,Nb)-(Cu,Ni,Co), (ii)
(Y,Ln)-Al-(Cu,Ni,Co), and (iii) (Au,Pd,Pt)-(Cu,Ni)-(Si,P).  (See 
Table~\ref{table:gfa} in Appendix~\ref{app1}.)


\section{Conclusion} 
\label{conclusion}

We performed event-driven molecular dynamics simulations of ternary
hard-sphere systems over the full range of compositions and diameter
ratios to identify the optimal glass-forming parameter regime.  We
identify two mechanisms for optimizing the glass-forming ability in
ternary systems. First, if the sizes of the three components are
similar, {\it i.e.} less than $10\%$ deviation in the diameters, the
ternary system behaves effectively as a binary system containing only
the largest and smallest particles. Second, if the diameter ratio of
the smallest to the largest particle is below the demixing threshold
$\sigma_C/\sigma_A \lesssim 0.8$, adding the third component $B$ with
$\sigma_C/\sigma_A < \sigma_B/\sigma_A < 1$ will dramatically enhance
the glass-forming ability above that for binary systems.  We show that
all experimentally observed ternary BMGs to date possess atomic
species for which the diameter ratio of the smallest to the largest
satisfies $\sigma_C/\sigma_A < 0.8$. Thus, an efficient strategy to design
BMGs with good GFA is to maintain large atomic size differences and 
prevent demixing by introducing three or more atomic components.

We recognize that the inter-atomic potentials describing BMGs are far
more complicated than the pairwise additive hard-sphere potential that
we employed. For example, the apparent distance between the repulsive
cores of two elements can be shorter than the mean core size of the
two elements~\cite{cheng:2009}. Also, in a more exact treatment,
Friedel oscillations originating from perturbations to the electron
density should be included since they are known to change the
stability of various crystalline lattices~\cite{doye:2003}.  Despite
these caveats and others, we show that the hard-sphere model is able
to semi-quantitatively predict the regime of optimal glass-forming
ability in experimentally observed ternary BMGs. In the near future,
we will consider how non-additivity of the particle diameters,
attractive interactions and barriers in the pairwise potential, and
multi-body interactions affect crystal and glass formation
and modify the hard-sphere predictions~\cite{zhang:2015}.

\begin{acknowledgments}
The authors acknowledge primary financial support from the NSF MRSEC
DMR-1119826 (KZ and BD) and partial support from NSF grant numbers
DMR-1006537 (CO) and CBET-0968013 (MS).  We also acknowledge support
from the Kavli Institute for Theoretical Physics (through NSF Grant
No.  PHY-1125915), where some of this work was performed. This work
also benefited from the facilities and staff of the Yale University
Faculty of Arts and Sciences High Performance Computing Center and the
NSF (Grant No. CNS-0821132) that in part funded acquisition of the
computational facilities.
\end{acknowledgments}

\begin{appendix}

\section{Glass-forming ability of experimentally observed BMGs}
\label{app1}

In this appendix, we provide a table of the critical cooling rates
$\mathcal{R}_c$ in units of ${\rm K/s}$ for experimentally observed binary and
ternary BMGs. See Table~\ref{table:gfa}. In the first column, we list
each class of binary and ternary BMGs according to the atom types that are 
present. Atom
types with similar sizes and properties are grouped together in
parentheses.  In the second column, we provide examples of specific
alloys within each BMG class.  The third column gives diameter ratios:
$\sigma_B/\sigma_A$ with $\sigma_B \le \sigma_A$ for binary systems and
$\sigma_B/\sigma_A$ and $\sigma_C/\sigma_A$ with $\sigma_C \le
\sigma_B \le \sigma_A$ for ternary
systems~\cite{egami:1984,inoue:2000,miracle:2003,long:2009}.

\begin{table*}
\begin{center}
\begin{threeparttable}
\caption{Glass-forming ability (characterized by the critical cooling 
rate $\mathcal{R}_c$ in units of ${\rm K/s}$) and atomic diameter ratio$^a$ for 
experimentally observed binary and ternary BMGs~\cite{inoue:2000,miracle:2003,long:2009}. Elements with similar sizes and properties are grouped together using parentheses (see also Fig.~\ref{fig:experiment}). The lanthanide elements$^b$ are indicated by Ln. }   
\begin{tabular}{c  c  c c}
\hline
Binary system &  Alloy  & $\mathcal{R}_c$~(K/s) & Diameter ratio $\sigma_B/\sigma_A$\\
\hline
Fe-B &${\rm Fe_{91}B_{9}}$ & $2.6\times 10^{7}$ &  0.62\\
(Au,Pd)-Si &  &  &  \\
 & ${\rm Au_{80}Si_{20}}$ & $1.0\times 10^{6}$ & 0.71 \\
&${\rm Pd_{95}Si_{5}}$ & $5.0\times 10^{7}$ &  0.74\\
&${\rm Pd_{82}Si_{18}}$ & $1.8\times 10^{3}$ & \\
&${\rm Pd_{75}Si_{25}}$ & $1.0\times 10^{6}$ & \\
Ti-Be&${\rm Ti_{63}Be_{37}}$ & $6.3\times 10^{6}$ & 0.76\\
Zr-Be&${\rm Zr_{65}Be_{35}}$ & $1.0\times 10^{7}$  & 0.7\\
Zr-Cu&${\rm Zr_{50}Cu_{50}}$ & 250 & 0.8\\
Nb-Ni&${\rm Nb_{40}Ni_{60}}$ &1400 & 0.85\\
\hline
Ternary system &  Alloy$^c$  & $\mathcal{R}_c$~(K/s) & Diameter ratios $\sigma_C/
\sigma_A$, $\sigma_B/\sigma_A$\\
\hline
Au-Si-Ge & ${\rm Au_{77.8}Si_{8.4}Ge_{13.8}}$ &  $3.0\times 10^{6}$ & 0.71, 0.79\\
(Au,Pd,Pt)-(Cu,Ni)-(Si,P) & & &  \\
&${\rm Pd_{40}Ni_{40}P_{20}}$ & 0.167 & 0.73, 0.91\\
&${\rm Pd_{77}Cu_{6}Si_{17}}$ & 125 & 0.74, 0.93\\
&${\rm Pd_{79.5}Cu_{4}Si_{16.5}}$ & 500 & \\
&${\rm Pd_{77.5}Cu_{6}Si_{16.5}}$ & 100 & \\
(Y,Ca,Ln)-Mg-(Cu,Ni) & & & \\
  & ${\rm Nd_{15}Mg_{70}Ni_{15}}$ & 178.2& 0.69, 0.88\\
  & ${\rm Nd_{15}Mg_{65}Ni_{20}}$ & 30&  \\
  & ${\rm Nd_{10}Mg_{75}Ni_{15}}$ & 46.1& \\
    & ${\rm Nd_{5}Mg_{77}Ni_{18}}$ & 49000& \\
   & ${\rm Nd_{5}Mg_{90}Ni_{5}}$ & 53000&  \\
     & ${\rm Nd_{10}Mg_{80}Ni_{10}}$ & 1251.4& \\
   & ${\rm Y_{10}Mg_{65}Cu_{25}}$ & 50 & 0.71, 0.89\\
 & ${\rm Gd_{10}Mg_{65}Cu_{25}}$ & 1 & 0.71, 0.89\\
 (Y,Ln)-Al-(Cu,Ni,Co)  &&  & \\
&${\rm La_{55}Al_{25}Ni_{20}}$ & 67.5 & 0.66, 0.76\\
&${\rm La_{55}Al_{25}Cu_{20}}$ & 72.3 & 0.68, 0.76\\
&${\rm La_{66}Al_{14}Cu_{20}}$ & 37.5 & \\
(Zr,Hf,Sn,Mg)-(Al,Ti,Nb)-(Cu,Ni,Co) & &  & \\
&${\rm Zr_{66}Al_{8}Ni_{26}}$ & 66.6 & 0.78, 0.89\\
(Zn,Al,Ag)-Mg-Ca & & & \\
&${\rm Zn_{20}Mg_{15}Ca_{65}}$ & 20 & 0.69, 0.81\\
\hline
\end{tabular}
\label{table:gfa}
\begin{tablenotes}
\item [a] Atomic radii are obtained from Refs.~\cite{egami:1984,miracle:2003}, which determine the atomic sizes using the first peak of the radial distribution function of amorphous liquid alloys or half of the spacing between 
atoms in metallic solids.
\item [b] Ln refers to the series of fifteen metallic elements (La, Ce, Pr, Nd, Pm, Sm, Eu, Gd, Tb, Dy, Ho, Er, Tm, Yb, and Lu) with atomic numbers $57$-$71$. Together with two more chemically similar elements Sc and Y, these seventeen elements are collectively known as the rare earth elements and are typically the largest sized component in BMGs.
\item [c] For each system, we only list alloys for which the critical cooling rate $\mathcal{R}_c$ has been reported. Other alloys such as Ca-Mg-Cu, Hf-Al-Cu, and Y-Al-Co, are also BMG formers, but with unreported values of $\mathcal{R}_c$.
\end{tablenotes}
\end{threeparttable}
\end{center}
\end{table*}

\section{Measurement of Critical Compression Rate $R_c$ in Simulations}
\label{app2}

\begin{figure}[H]
\includegraphics[width=3.5in,center]{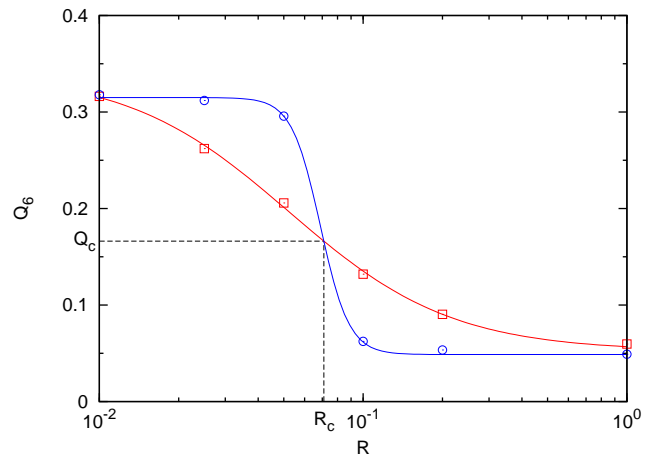}

\caption{(Color online) The mean and median $Q_6$ as a function of
compression rate $R$ for ternary hard spheres with diameter ratios
$\sigma_C/\sigma_A=0.88$ and $\sigma_B/\sigma_A=0.9$ and compositions
$x_A=x_B=x_C=1/3$. The critical compression rate $R_c$ and bond orientational 
order parameter $Q_c$ are defined by the intersection of the mean and 
median $Q_6$. The solid lines are fits to a logistic function. }
\label{fig:q6}
\end{figure}

In this appendix, we describe the measurement of the critical
compression rate $R_c$ in the molecular dynamics simulations of
hard-spheres. We performed $96$ independent runs to generate jammed
configurations at each compression rate $R$. We find that the
distribution $P(Q_6)$ of the global bond orientational order parameter
becomes bimodal as $R\to R_c$ with peaks that
correspond to amorphous and partially crystalline configurations.  In
Fig.~\ref{fig:q6}, we show that both the mean and median $Q_6$ possess
a sigmoidal shape on a logarithmic scale in $R$.  We define $R_c$ as
the critical compression rate at which the mean and median $Q_6$
intersect. A configuration is determined to be crystalline (amorphous)
if $Q_6 > Q_c$ ($Q_6 < Q_c$), where $Q_c$ is the value of $Q_6$ at
which the mean and median $Q_6$ intersect.  Since the distribution
$P(Q_6)$ is bimodal at $R \approx R_c$, one can also fit $P(Q_6)$
by the sum of two Gaussian distributions and identify $R_c$ as the rate
at which the two Gaussian contributions have equal area. This method can 
be more time efficient since it avoids
measurements at $R \ll R_c$.

\end{appendix}


\end{document}